\documentclass[
    ,final            % use final for the camera ready runs
  ]
  {aipproc}

\layoutstyle{6x9}

\begin{document}

\title{Dark Matter tested with satellites}


\author{F. Combes}{
  address={Observatoire de Paris, LERMA, 61 Av. de l'Observatoire, F-75014 Paris, France}
}

\author{O. Tiret}{
  address={SISSA, via Beirut 4, I-34014 Trieste, Italy}
}

\begin{abstract}
Recently, the distribution of velocity dispersion as far as 400kpc around
red isolated galaxies was derived from statistical studies of satellites
in the SDSS (Klypin \& Prada 2009). This could help to constrain dark
matter models at intermediate scales. We compare the predictions of
different DM distributions, $\Lambda$CDM with NFW or cored 
profiles, and also modified
gravity models, with observations. It is shown how the freedom in the
various parameters (radial distribution of satellites, velocity
anisotropy, external field effect), prevents to disentangle the models,
which all can give pretty good fits to the data. In all cases, realistic 
radial variations of velocity anisotropy are used for the satellites, 
and a constant stellar-mass to light ratio for the host galaxies.
\end{abstract}

\maketitle

%%%%%%%%%%%%%%%%%%%%%%%%%%%%%%%%%%%%%%%%%%%%
%% MAINMATTER
%%%%%%%%%%%%%%%%%%%%%%%%%%%%%%%%%%%%%%%%%%%%

\section{Method of satellite kinematics}

To obtain the dark matter distribution at large scale around galaxies,
much farther than the extent of rotation curves, astronomers
have since a long time used the kinematics of satellite galaxies
(Erickson et al 1987, Zaritsky et al  1993, 1994, 1997).
Because the number of satellites per galaxy is very small, the technique
consists in stacking the data on many galaxy-satellite pairs, in common
bins of host luminosity. In the 1990s, the number of satellites were 
counted in hundreds, and the statistics were not sufficient to provide
significant results. For instance with 
 115 satellites at distances lower than  500kpc, the probability to
find Nsat satellites was fitted to P(Nsat) = 0.4$^{Nsat}$ 
(Zaritsky et al 1997). No dependence of velocity dispersion
of the satellites with radius, nor with the host mass, was found.

The advent of rich surveys such as the SDSS and 2dF GRS
have given more statistical value to the satellite studies, including now
Nsat = a few thousands (McKay et al 2002, Brainerd \& Specian 2003,  
Prada et al 2003, van den Bosch et al 2004).
While some studies still find a velocity dispersion $\sigma$  flat with radius,
most of them now see the decrease of $\sigma$  with radius, and 
the increase with the host luminosity or mass.  However, there are
still contradictory results, as reviewed by Norberg et al (2008).
 They found that the velocity dispersion within a projected radius of
175$h^{-1}$kpc is increasing as the square root of luminosity 
for ellipticals, implying a constant M/L ratio with mass.
For spirals the slope is higher, $\sigma_v \propto L^{0.8}$
and the mass-to-light ratio is increasing with luminosity.
M/L is 3-10 times higher for ellipticals than for spirals,
at the same luminosity L$_{g}$.
For the Milky Way luminosity bin, M(175$h^{-1}$kpc)=
$3.5 10^{11} h^{-1} M_\odot$, but there is a 
large scatter, for a given L.

\subsection{Caveats of the satellite method}

One of the main problem of the method is the stacking of satellites
in host luminosity bins, to recover sufficient statistics.
In each L bin, there is a significant scatter in halo mass,
which prevents to derive a unique halo mass-luminosity relation
(More et al 2009a).
To help raise the degeneracy, it is useful to compute
the velocity dispersion by different weighting methods:
the satellite weighted dispersion:
$\sigma_{sw}^2 = 1/N_{sat} \Sigma N_j \sigma_j^2$, 
(where the sum is over the $N_c$ central galaxies, identified 
by their $j^{th}$ number, who have each $N_j$ satellites),
is biased towards higher values than the  host weighted one:
$\sigma_{hw}^2 = 1/N_{c} \Sigma \sigma_j^2$,
and the scatter increases with luminosity (More et al 2009b).

One other caveat is to discriminate against interlopers, 
the fraction of which depends on the host selection criteria;
the interlopers fraction has been estimated from 10\%
for extremely isolated hosts (Prada et al 2003), 
to 30\% or more,  in mock catalogs (van den Bosch et al 2004).
The difficulty is that interlopers are not uniformly distributed
on the sky, but are also clustered (e.g. Chen et al 2006). 

\subsection{Tully-Fisher equivalent}

With the help of the mock simulations to interpret the data,
a quiterobust result is that halo mass to luminosity M/L decreases with L, and 
more specifically, the dispersion-luminosity relation depends
on the radius it is estimated:
$\sigma_v \propto L^{0.3}$ at 120kpc and 
$\sigma_v \propto L^{0.5}$  at 350kpc (Prada et al 2003).

It is interesting to compare the results with those obtained 
through weak lensing (Hoekstra et al 2002). The slope of the 
dispersion-luminosity relation is very close to the Tully-Fisher slope
1/4 for spiral galaxies (Verheijen 2001), and can be called
a Tully-Fisher equivalent.

\subsection{Velocity anisotropy}

Recently, Klypin \& Prada (2009) have carried out
a further satellite study from the SDSS, selecting as hosts only
red isolated galaxies, expected to be ellipticals or spheroids.
Since the hosts are very isolated, there are 
only 1 or 0 satellite for each galaxy. They compute
the radial distribution of velocity dispersion, in three host luminosity
bins, and find constraints on dark matter models 
($\Lambda$CDM or modified gravity MOND), while
fitting both the dispersion $\sigma$, and the radial density law of the satellites.
They claim that MOND cannot account for the observations.

The fits however have to include many free parameters, 
related to the velocity anisotropy of satellites, and therefore
contain significant degeneracy. The situation is quite similar
to the velocity dispersion studies at smaller scales around
elliptical galaxies. The drop of $\sigma$ derived from
planetary nebulae was first interpreted as a possible
dearth of dark matter (Romanowsky et al 2003), while
fits including a radially variable  anisotropy of velocities
$\beta$ reconciled the data with the CDM model (Dekel et al 2005).
$\beta$ can vary between -$\infty$ (purely tangential orbits),
to 1 (purely radial orbits), passing through 0 (isotropy),
and this is justified through galaxy mergers, since ellipticals
are assumed to be the result of major mergers, or a succession
of minor mergers.

\section{Tests of the gravity with the SDSS satellites}

We used the satellite kinematics data from Klypin
\& Prada (2009) to test the predictions of MOND at
large-scale around isolated galaxies. It has been
shown that when the anisotropy parameter radial variation
is properly taken into account, MOND gives a good fit
of the data (Angus et al 2007, Tiret et al 2007).
We also combined around a typical well-studied  early-type galaxy
NGC 3379 the various fits of velocity dispersion at three different scales:
small-scale with stellar tracers, intermediate scale with planetary nebulae,
and large-scale with satellites (in the corresponding luminosity bin).
Both CDM and MOND models provide satisfying fits (although
CDM cores have to be assumed in the center, Tiret et al 2007).
The anisotropy is comparable to what is expected in 
cosmological simulations (Sales et al 2007).

\begin{figure}
  \includegraphics[angle=-90,width=.9\textwidth]{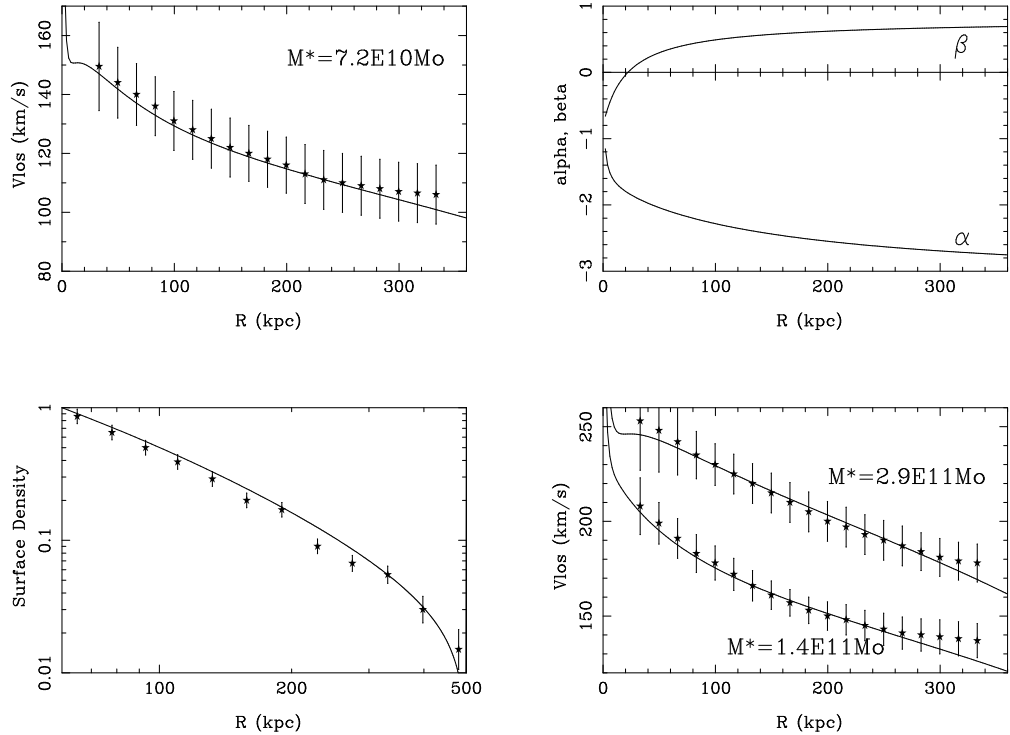}
  \caption{Fit with MOND of the line of sight rms velocities of galaxies,
with the stellar mass indicated in the plots, taken from Klypin \& Prada (2009).
The upper right panel displays the slope $\alpha$ of the tracer density
($\alpha = dlog \rho / dlog r$) and the anisotropy
 $\beta= 1 - \sigma_{\theta}^2 /\sigma_r^2$
of velocity used in the fit of the M* = 7.2 10$^{10}$ M$_\odot$
case, and the bottom left panel is a fit of the tracer surface density, for the same case.}
\end{figure}

We present in Figure 1 more fits, corresponding to
different mass and luminosity intervals, not available before,
together with the radial distribution of tracer density and velocity anisotropy.  
 For all these fits, a constant stellar mass to light ratio of $M/L_g$=4 has
been adopted, corresponding to realistic populations for the red galaxies
involved.  Note that given the Tully-Fisher equivalent relation observed,
i.e. $\sigma_v \propto L^{0.25}$, the fit can easily be generalised 
for a wide range of mass, since these outside regions are in the MOND regime, 
where  $\sigma_v^2 \propto \sqrt{a_0 M}$ (Milgrom 1983).

These fits are for isolated galaxies. We might expect problems in groups and clusters,
when large masses exist nearby. They produce then an External Field Effect (EFE),
that reduce the dark matter equivalent of MOND (although the EFE
has been successfully fitted in the Milky Way, Wu et al 2007).
At even larger scales, the dark halo mass-to-luminosity 
 depends on environment, and for groups on the
crossing time.
Small and large haloes have the largest M/L, while
intermediate haloes, small groups with late-type galaxies
have the lowest M/L of $ \sim$90  (Tully 2005).

\section{Conclusion}

The method of satellite kinematics is giving
now more robust results, with the increased statistics of
big surveys (SDSS, 2dF).  However,  still very different results 
can be found in the literature, according to the selection of
primaries (isolation criterium), and the elimination of interlopers.
The  mass and radial dependence of $\sigma_v$  is now
derived, but with large uncertainties, due essentially to stacking problems,
the mass being widely scattered in a given luminosity bin.

The modelisation involves numerous degrees of freedom, in the radial
distribution of the velocity anistropy, essentially. 
The shape of $\sigma_v$  versus distance can be fit with appropriate $\beta$
both in $\Lambda$CDM and MOND. The generalisation to a large
range of masses, is automatic if the Tully-Fisher equivalent relation is satisfied 
$\sigma_v \propto L^{0.25}$.

%%%%%%%%%%%%%%%%%%%%%%%%%%%%%%%%%%%%%%%%%%%%%%%%
%% BACKMATTER
%%%%%%%%%%%%%%%%%%%%%%%%%%%%%%%%%%%%%%%%%%%%%%%%

%\begin{theacknowledgments}
%\end{theacknowledgments}


%%%%%%%%%%%%%%%%%%%%%%%%%%%%%%%%%%%%%%%%%%%
%% The following lines show an example how to produce a bibliography
%% without the help of the BibTeX program. This could be used instead
%% of the above.
%%%%%%%%%%%%%%%%%%%%%%%%%%%%%%%%%%%%%%%%%%%

%%
\end{document}